\documentclass[twocolumn,showpacs,preprintnumbers,amsmath,amssymb,superscriptaddress]{revtex4}

\usepackage{graphicx}  
\usepackage{dcolumn}   
\usepackage{bm}        
\usepackage{stmaryrd}  

\usepackage{color}

\def  \btau    {\bm\tau}
\def  \be      {\hat{\bm e}}
\def  \bhats   {\hat{\bm s}}
\def  \bhatS   {\hat{\bm S}}
\def  \d       {{\rm d}}

\begin{document}

\preprint{APS/123-QED}

\title{Current-driven destabilization of both collinear configurations
in asymmetric spin-valves}

\author{M. Gmitra}
\affiliation{Department of Physics, Adam Mickiewicz University,
             Umultowska 85, 61-614~Pozna\'n, Poland}
\author{J. Barna\'s}
\affiliation{Department of Physics, Adam Mickiewicz University,
             Umultowska 85, 61-614~Pozna\'n, Poland}
\affiliation{Institute of Molecular Physics, Polish Academy of Sciences,
             M.~Smoluchowskiego~17, 60-179~Pozna\'n, Poland}

\date{\today}

\begin{abstract}
Spin transfer torque in spin valves usually destabilizes one of
the collinear configurations (either parallel or antiparallel) and
stabilizes the second one. Apart from this, balance of the
spin-transfer and damping torques can lead to steady precessional
modes. In this letter we show that in some asymmetric nanopillars
spin current can destabilize both parallel and antiparallel
configurations. As a result, stationary precessional modes can
occur at zero magnetic field. The corresponding phase diagram as
well as frequencies of the precessional modes have been calculated
in the framework of macrospin model. The relevant spin transfer
torque has been calculated in terms of the macroscopic model based
on spin diffusion equations. 
\end{abstract}

 \maketitle

{\it Introduction:} Spin transfer between conduction electron
system and localized magnetic moments gives rise to new phenomena,
like for instance current-induced switching between different
magnetic states \cite{Slonczewski96}. Several theoretical models
have been developed to describe physical mechanisms of the spin
transfer and magnetic switching phenomena
\cite{Stiles02,Brataas01,weintal00,zhang02,Barnas05}. Apart from
this, magnetic switching has been observed in a number of
experiments \cite{Katine00,grolier01,darwish04,tsoi04}.

At certain conditions electric current can cause transition to
steady precessional modes, where the energy is pumped from
conduction electrons to localized magnetic moments
\cite{Slavin05,Sun00Li03,Kiselev0305,Rippard04,Xi04,Xiao05}. This
phenomenon is of high importance due to possible applications in
microwave generation. In typical Co/Cu/Co spin valves, the steady
precessions exist for external magnetic fields larger than the
anisotropy field, and for currents exceeding certain critical
values \cite{Kiselev0305,Rippard04}. For lower values of external
field, the current drives switching to antiparallel (AP) or
parallel (P) states, depending on the initial state of the system.

In this Letter we show that spin transfer torque in asymmetric
systems (with the two magnetic films having different bulk and
interface spin asymmetry factors, e.g., Co/Cu/Py nanopillars)
vanishes at a certain noncollinear configuration due to an inverse
spin accumulation in the nonmagnetic spacer layer. As a result,
spin current can destabilize both P and AP magnetic configurations
for one orientation of the bias current and stabilize both
configurations for the opposite current. The former case is of
particular interest, as the spin current can excite precessional
modes in the absence of external magnetic field, which is of
particular importance from the application point of view. Such a
system is thoroughly studied in this Letter.

{\it Model and description:} We consider a tri-layer system which
is attached to two external nonmagnetic leads. The tri-layer
consists of a magnetically fixed (thick) layer of thickness $d_1$,
nonmagnetic spacer layer of thickness $d_2$, and a thin magnetic
(free or sensing) layer of thickness $d_3$. Magnetic dynamics of
the sensing layer is described by the generalized
Landau-Lifshitz-Gilbert equation
\begin{equation}
\label{Eq:LLG}
\frac{\d\bhats}{\d t} = - |\gamma_{\rm g}|\mu_0\, \bhats \times {\bm H}_{\rm eff}
- \alpha\,\bhats \times \frac{\d\bhats}{\d t}
+ \frac{|\gamma_{\rm g}|}{ M_{\rm s} d_3} \, \btau \,,
\end{equation}
where $\bhats$ is the unit vector along the spin moment of the
sensing layer, $\gamma_{\rm g}$ is the gyromagnetic ratio, $\mu_0$
is the magnetic vacuum permeability, ${\bm H}_{\rm eff}$ is an
effective magnetic field acting on the sensing layer, $\alpha$ is
the damping parameter, and $M_{\rm s}$ stands for the saturation
magnetization. Generally, the effective field includes an external
magnetic field $H_{\rm ext}$, the uniaxial magnetic anisotropy
field $H_{\rm a}$, and the demagnetization field $H_{\rm d}$;
${\bm H}_{\rm eff} = - H_{\rm ext}\,\be_{\rm z} - H_{\rm a} (
\bhats\cdot\be_z ) \, \be_z + H_{\rm d} (\bhats\cdot\be_x )\,
\be_x$, where $\be_x$ and $\be_z$ are the unit vectors along the
axes $x$ (normal to the layers) and $z$ (in-plane), respectively.
The last term in Eq.(\ref{Eq:LLG}) stands for the torque due to
spin transfer, $\btau = \btau_\theta + \btau_\varphi$, with
$\btau_\theta = aI\, \bhats\times(\bhats\times\bhatS) =
\tau_\theta \be_\theta$, and $\btau_\varphi = bI\,
\bhats\times\bhatS =\tau_\varphi \be_\varphi$. Here, $\bhatS$ is
the unit vector along the spin moment of the fixed magnetic layer
($\bhatS=\be_z$), while $\be_\theta$ and $\be_\varphi$ are the
unit vectors of a coordinate system associated with the polar
$\theta$ and azimuthal $\varphi$ angles describing orientation of
the vector $\bhats$. The current $I$ is defined as positive when
it flows from the sensing layer towards the thick magnetic one
(opposite to the axis $x$).

The current-induced torque $\btau$ and the corresponding
parameters $a$ and $b$ have been calculated in the diffusive
transport regime, assuming that the spin current component
perpendicular to the spin moment of the sensing layer is entirely
absorbed in the interfacial region
\cite{Stiles02,Brataas01,Barnas05}. For the parameter $a$ one then
finds \cite{Barnas05}
\begin{eqnarray}
a=\frac{\hbar}{e^2}\bigg[ {\rm Re}\{G_{\uparrow\downarrow}\}\bigg(
\cot\theta(\tilde{g}_x\cos\varphi + \tilde{g}_y\sin\varphi) -
\tilde{g}_z\bigg)
\nonumber \\
+ \frac{{\rm Im}\{G_{\uparrow\downarrow}\}}{\sin\theta}\bigg(
\tilde{g}_x\sin\varphi -  \tilde{g}_y\cos\varphi\bigg)
\bigg]\bigg|_{x\to x_0^-}\,,\,\,\,
\end{eqnarray}
where $\tilde{\bf g}={\bf g}/I$, with the spin accumulation
components ${\bf g}=(g_x, g_y, g_z)$ taken in the nonmagnetic
spacer at the very interface with the sensing layer $(x\to
x_0^-)$, and $G_{\uparrow\downarrow}$ is the mixing conductance
\cite{Brataas01}. The parameter $b$ is given by a similar formula,
but with ${\rm Re}\{G_{\uparrow\downarrow}\}$ replaced by $-{\rm
Im}\{G_{\uparrow\downarrow}\}$ and ${\rm
Im}\{G_{\uparrow\downarrow}\}$ replaced by ${\rm
Re}\{G_{\uparrow\downarrow}\}$.

A qualitative picture of the non-linear dynamics can be obtained
from the local phase portraits of the corresponding linearized
system in the vicinity of fixed points. Accordingly, we linearized
Eqs.(\ref{Eq:LLG}) and calculated eigenvalues of the corresponding
Jacobian. For a two dimensional problem (as in our case) the
eigenvalues depend only on the trace and determinant of the
Jacobian. Apart from this, the eigenvalues are current dependent
and determine stability of the fixed points. More specifically,
the point is stable when real parts of all the eigenvalues are
negative, and becomes unstable when at least one of them is
positive. The system considered, Eqs.(\ref{Eq:LLG}), can have
several fixed points given by $\btau=\mu_0 M_{\rm s} d_3
\bhats\times{\bm H}_{\rm eff}$. However, only two of them, P and
AP, are trivial with position independent of current. In addition,
the current driven global phase portrait of the system considered
below contains two saddle points with separatrices dividing the
phase space to basin of attraction for P and AP fixed points and
two focuses (located near energy maximum).

{\it Critical currents:} Consider first the P state and assume
positive determinant of the corresponding Jacobian. Since the
eigenvalues of the linearized problem can be complex numbers, the
non-zero imaginary parts give rise to periodic components of the
fundamental solutions. The condition of vanishing trace determines
the critical current $I_{\rm c}^{\rm PRC}$ which destabilizes the
P state and switches system to a precessional (PRC) state (this
can be assign as Hopf bifurcation with a limit cycle emerging),
\begin{equation}
\label{Eq:ic1} I_{\rm c}^{\rm PRC}=\frac{\alpha\mu_0 M_{\rm s} d_3
}{a-b\,\alpha} \left( H_{\rm a}+H_{\rm ext}+\frac{H_{\rm
d}}{2}\right) \bigg|_{\theta\to 0}\,,
\end{equation}
where the parameters $a$ and $b$ have to be calculated in the
limit of $\theta\to 0$. With a further increase in current, the
complex conjugated imaginary parts of the eigenvalues vanish at a
certain point, at which two real eigenvalue branches arise. When
current increases further, one of the eigenvalues goes to zero,
leading to a saddle-node bifurcation.  In the case studied below,
two new saddles appear inside the limit cycle making it unstable
\cite{wiggins}. Thus, the critical currents driving system to a
static state (SS) can be found as,
\begin{eqnarray}
\label{Eq:ic2} I_{{\rm c}\pm}^{\rm SS}=-\frac{\mu_0 M_{\rm s} d_3
}{a^2+b^2}\bigg[ b\left( H_{\rm a}+H_{\rm ext}+\frac{H_{\rm
d}}{2}\right) \qquad \\ \nonumber \pm \frac{1}{2}\sqrt{b^2 H_{\rm
d}^2-4a^2(H_{\rm a}+H_{\rm ext})(H_{\rm a}+H_{\rm ext}+H_{\rm d})}
\bigg]\bigg|_{\theta\to 0}\,.
\end{eqnarray}

The signs of critical currents depend on the parameters $a$ and
$b$ taken at $\theta\rightarrow 0$. In the case considered below,
$I>0$, these parameters obey the conditions $a-b\alpha >0$ and
$b<0$ (the following discussion is limited to the case, where
these conditions are fulfilled). Consequently, the PRC regime
holds for $I>I_{\rm c}^{\rm PRC}$, and the P state is stable for
$I<I_{\rm c}^{\rm PRC}$. Since $I_{{\rm c}+}^{\rm SS}<I_{{\rm
c}-}^{\rm SS}$, the SS states can occur for $I>I_{{\rm c}+}^{\rm
SS}$, and we can skip $I_{{\rm c}-}^{\rm SS}$ in the following
considerations.

{\it Asymmetric nanopillars:} Let us consider now an asymmetric
nanopillar, where the two ferromagnets and/or two external leads
are of different materials, like for instance Cu/Co/Cu/Py/Cu spin
valve. It has been predicted recently that the spin-transfer
torque acting on the sensing layer in such a spin valve can vanish
for a noncollinear magnetic configuration \cite{Barnas05}. In
Fig.1(b,c) we plot the angular variation of the torques
$\tau_\varphi$ and $\tau_\theta$, which clearly shows that
$\tau_\theta$ vanishes at $\theta_c\ne 0,\pi$. Such a behavior of
$\tau_\theta$ can be observed if an inverse spin accumulation
builds up in the spacer layer, as shown schematically in Fig.1(a).
This can be achieved when the fixed magnetic layer has bulk spin
asymmetry weaker than that of the sensing layer, and its thickness
is smaller than the spin diffusion length.
\begin{figure}[!h]
\includegraphics[width=0.99\columnwidth]{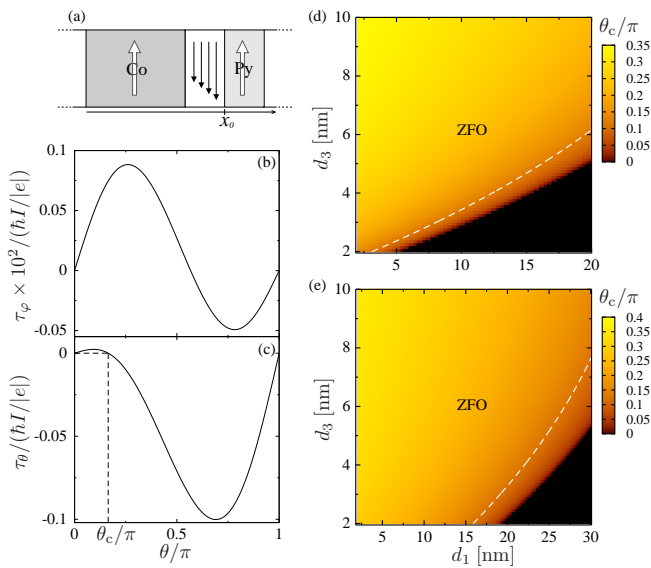}
\caption{(color online) (a)~Schematic structure of the nanopillar
and spin accumulation in thin spacer layer for the parallel
configuration of the magnetic layers, and for electrons flowing
along the axis $x$. Angular dependence of the torques
$\tau_\varphi$~(b) and $\tau_\theta$~(c), acting on the Py layer
in the Cu/Co(10)/Cu(10)/Py(4)/Cu spin valve (independent of
$\varphi$). (d)~$\theta_{\rm c}/\pi$ as a function of $d_1$ and
$d_3$ in the nanopillar Cu/Co($d_1$)/Cu(10)/Py($d_3$)/Cu. The
white dashed line corresponds to a critical thicknesses for ZFOs
(zero-field oscillations). (e)~$\theta_{\rm c}/\pi$ in the
Cu/Co($d_1$)/Cu(10)/Py($d_3$)/Au structure. The other parameters
are: (Co) bulk resistivity $\rho^*=5.1\,{\rm\mu\Omega cm}$, spin
asymmetry factor $\beta=0.51$, and spin-flip length $l_{\rm
sf}=60\,{\rm nm}$; (Py) $\rho^*=16\,{\rm\mu\Omega cm}$,
$\beta=0.77$,  $l_{\rm sf}=5.5\,{\rm nm}$; (Cu)
$\rho^*=0.5\,{\rm\mu\Omega cm}$, $l_{\rm sf}=1000\,{\rm nm}$; (Au)
$\rho^*=2\,{\rm\mu\Omega cm}$ and $l_{\rm sf}=60\,{\rm nm}$. In
turn, for the Co/Cu interfaces we assume the interfacial
resistance per unit square $R^*=0.52\times 10^{-15}\,{\rm\Omega
m^2}$, interface spin asymmetry factor $\gamma=0.76$, and the
mixing conductances ${\rm
Re}\{G_{\uparrow\downarrow}\}=0.542\times
10^{15}\,{\rm\Omega^{-1}m^{-2}}$, and ${\rm
Im}\{G_{\uparrow\downarrow}\}=0.016\times
10^{15}\,{\rm\Omega^{-1}m^{-2}}$; for the Py/Cu interfaces we
assume $R^*=0.5\times 10^{-15}\,{\rm\Omega m^2}$, $\gamma=0.7$,
${\rm Re}\{G_{\uparrow\downarrow}\}=0.39\times
10^{15}\,{\rm\Omega^{-1}m^{-2}}$, and ${\rm
Im}\{G_{\uparrow\downarrow}\}=0.012\times
10^{15}\,{\rm\Omega^{-1}m^{-2}}$. }
\end{figure}
The general conditions for vanishing spin torque at a noncollinear
configuration can be derived assuming $\partial\tau_\theta
/\partial\theta > 0$ in both P and AP states, and for in-plane
configurations ($\varphi\to\pi/2$) it has the form
\begin{equation}
\label{Eq:cond1}
\left( \pm{\tilde g}_z - \frac{{\rm Im}\{G_{\uparrow\downarrow}\}}
                              {{\rm Re}\{G_{\uparrow\downarrow}\}}
       \frac{\partial {\tilde g}_x}{\partial\theta} \mp
       \frac{\partial {\tilde g}_y}{\partial\theta}
\right)
\Bigg|_{\scriptsize\begin{array}{l}\theta\to 0 \\[-2.0pt] \theta\to\pi \end{array}
} < 0 \,.
\end{equation}
For a general configuration one has to replace ${\tilde g}_x
\mapsto ({\tilde g}_x\sin\varphi-{\tilde g}_y\cos\varphi)$ and
${\tilde g}_y \mapsto ({\tilde g}_x\cos\varphi+{\tilde
g}_y\sin\varphi)$.

The parameter $\theta_{\rm c}/\pi$ is plotted in Fig.1 as a
function of $d_1$ and $d_3$ for the right lead made of Cu (d) and
Au (e), respectively. As one can note, $\theta_{\rm c}$ becomes
particularly large for thin Co layers and thick Py ones. However,
for a too large ratio $d_3/d_1$ the roles of the Co layer as a
fixed one and Py as a sensing layer could be interchanged.
Therefore, in the following we will consider the
Cu/Co(10)/Cu(10)/Py(4)/Cu nanopillar (the numbers are the layer
thicknesses in nanometers), with Py as a sensing layer and Co as
the magnetically fixed one. For the Py layer we assume $M_{\rm
s}=10053.1\,{\rm Oe}$, $H_{\rm a}=2.51\,{\rm Oe}$, $H_{\rm d}=0.65
M_{\rm s}$ and $\alpha=0.003$. The dependence of $\theta_{\rm c}$
on the spacer thickness is rather weak, so thicker spacers can be
used to eliminate possible interlayer exchange interaction. From
comparison of Figs 1(d) and 1(e) follows that reducing the
spin-flip length in the lead adjacent to the sensing layer
increases $\theta_{\rm c}$ and also enhances the torque acting on
the free layer (reducing the relevant critical current). This is
due to an increased spin accumulation at the Cu/Py interface. The
$\tau_\varphi$ component is usually much smaller than
$\tau_\theta$, and therefore plays a negligible role in the
initial stage of the switching from P state.

From the above follows that in asymmetric spin valves both P and
AP states can be unstable when current density exceeds a certain
critical value. Accordingly, precessional states and/or
bistability of the static states can be expected for certain
values of spin current \cite{Slavin05, Sun00Li03, Xi04}. If energy
of a static state is close to the magnetic energy maximum, and a
critical current for its stabilization is larger than the critical
current needed to destabilize the state corresponding to the
energy minimum, the PRC state is allowed \cite{Slavin05}. Thus,
the sufficient condition for the presence of oscillations reads
$I_{\rm c}^{\rm SS} > I_{\rm c}^{\rm PRC}$. The above condition,
together with the necessary condition of vanishing spin torque in
a noncollinear configuration, allow us to determine the region
where the zero-field oscillations (ZFOs) exist. In Fig.1(d-e), the
white dashed lines (corresponding to $I_{\rm c}^{\rm SS}=I_{\rm
c}^{\rm PRC}$) separate the regions where the ZFOs occur from
those where they cannot exist. We note that critical currents for
switching from the AP state are still given by Eqs (\ref{Eq:ic1})
and (\ref{Eq:ic2}), but with r.h.s. multiplied by $-1$ and $H_{\rm
ext}$ replaced by $-H_{\rm ext}$.

Figure 2 shows dynamical phase diagram of the Cu/Co/Cu/Py/Cu
nanopillar. The reduced magnetoresistance,
$r=(1-\cos^2(\theta/2))/(1+\cos^2(\theta/2))$, is shown there as a
function of the external field and reduced current density
$I/I_0$, with  $I_0=10^8\,{\rm A cm^{-2}}$. The dashed lines
correspond to the critical currents given by the analytical
formulas (\ref{Eq:ic1}) and (\ref{Eq:ic2}), and fit very well to
the numerical results. For weak external fields we find a stable
static state with a high magnetoresistance value (HSS), which is
close to maximal magnetic energy.
\begin{figure}[!h]
\includegraphics[width=0.999\columnwidth]{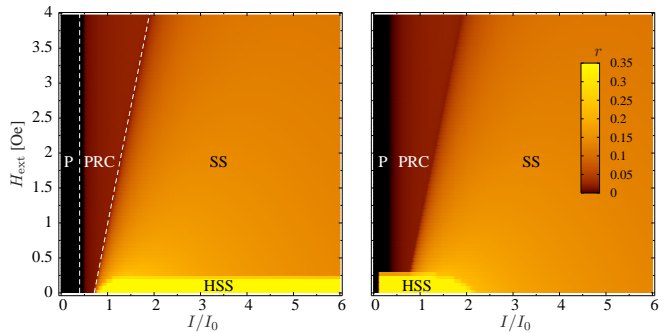}
\caption{(color online) Dynamical phase diagram of the
magnetoresistance for the Cu/Co(10)/Cu(10)/Py(4)/Cu spin valve,
scanned by increasing (left) and decreasing (right) current with
the sweeping rate $59$ A/cm$^{2}$s in a constant applied field.
The dashed lines correspond to critical currents given by Eqs
(\ref{Eq:ic1}) and (\ref{Eq:ic2}). The other parameters as in
Fig.1.}
\end{figure}
\begin{figure}[!h]
\includegraphics[width=0.8\columnwidth]{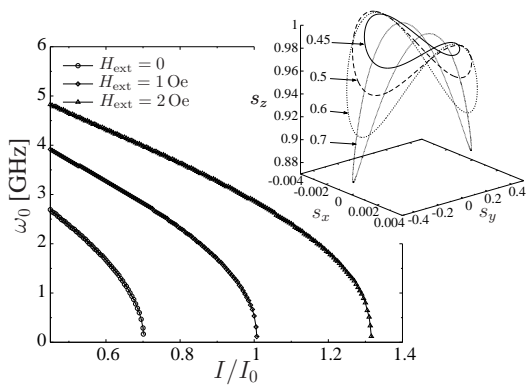}
\caption{Fundamental frequency of the magnetoresistance
oscillations in Cu/Co(10)/Cu(10)/Py(4)/Cu, calculated as a
function of the reduced current swept with the rate $16$
A/cm$^{2}$s in a constant external field as indicated. The inset
shows precession orbits in zero external magnetic field for
$I/I_0=0.45, 0.5, 0.6, 0.7$. The other parameters as in Fig.1. }
\end{figure}
The microwave oscillations exist in the area between the two
dashed white lines in Fig.2.(left part). As one can see, these
oscillations exist in the absence of magnetic field for a certain
current region. The fundamental frequency $\omega_0$ (first
harmonics) of the magnetoresistance oscillations is shown in Fig.3
as a function of the current density and for different magnetic
fields. Along the segments, where $\bhats$ moves almost in the
layer plane, the spin precesses mainly around $H_{\rm a}$ and
$H_{\rm ext}$, with the angular velocity proportional to
$|\gamma_{\rm g}| (H_{\rm a}+H_{\rm ext})$, see the inset in
Fig.3. Along the remaining part ($\bhats$ moves almost
perpendicularly to the layer plane), the angular velocity is
larger and proportional to $|\gamma_{\rm g}| (H_{\rm a}+ H_{\rm
ext}+H_{\rm d})$. With increasing $I$, the average orbital speed
decreases while the arc length of the orbit increases.
Consequently, $\omega_0$ decreases with increasing $I$.
\begin{figure}[!h]
\includegraphics[width=\columnwidth]{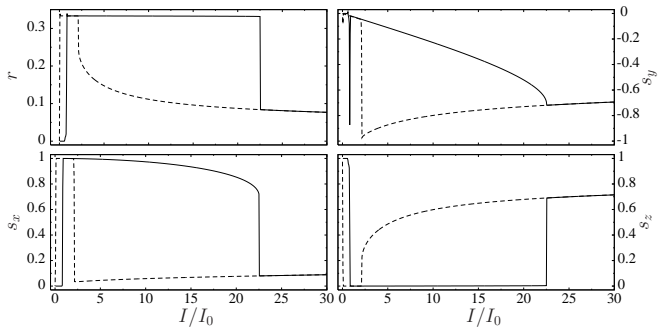}
\caption{Current driven hysteretic behavior of the
magnetoresistance and spin components of the sensing layer in zero
field, scanned with increasing (solid line) and decreasing (dashed
line) current with the sweeping rate $140$ A/cm$^{2}$s. The
parameters as in Fig.2. }
\end{figure}
Such precessions of decreasing frequency with increasing current
in symmetrical spin valves are called
'clamshell'\cite{Kiselev0305} or 'in-plane'\cite{Xiao05} modes.
But the main difference between symmetrical pillars and those
studied in this Letter concerns the way the spin transfer acts on
$\bhats$. In our case, increasing current leads to an increase in
$\btau$, and the orbits become narrower and elongated in the $z$
direction due to $\btau_\theta$. At the critical current density
given by Eq.(\ref{Eq:ic2}), when $s_z = \cos\theta_{\rm c}$,
$\tau_\theta$ vanishes and orbital velocity reaches minimum. The
damping torque is small, and the orbit bifurcates into SS states
(fixed points) due to a non-zero $\tau_\varphi$. A further current
increase at low fields drives the system {\it via} a transient
regime to HSS state with positive or negative $s_x$ component (see
Fig.4). The presence of HSS states gives rise to a current-driven
hysteresis (Figs.2 and 4). The hysteresis for lower fields can be
formally written as a sequence of transitions
P$\to$PRC$\to$SS$\to$${\rm T_+}$$\to$HSS$\to$${\rm T_-}$$\to$SS
for increasing current and SS$\to$${\rm T_+}$$\to$HSS$\to$${\rm
T_-}$$\to$OOP$\to$P/AP for the decreasing current, with final
bistability reached {\it via} the 'out-of-plane' precessions
(OOP), observed also in symmetrical spin-valves \cite{Xiao05}. The
switching from SS state to a HSS state is mediated by the
transient regime ${\rm T_+}$ (where spin transfer pumps energy to
the system). The HSS states can be observed in a quite broad
region, up to $I/I_0\simeq 22.5$, and further current increase
switches system {\it via} the next transient ${\rm T_-}$ regime
(where spin transfer dissipates energy) to the SS state. For
higher fields the HSS states become unstable and the hysteresis
vanishes. Finally we note that for $\tau_\varphi = 0$ the static
states will not occur and the area of PRC states extends for
$I>I_{\rm c}^{\rm PRC}$.

In conclusion, we have shown that spin polarized current in
certain asymmetrical structures can destabilize both P and AP
configurations of a nanopillar spin-valve, and drive microwave
oscillations in the absence of magnetic field.  This takes place
only for one orientation of the bias current. We have also
presented the corresponding  dynamical phase diagram.

{\it Acknowledgements} We thank A. Fert and D. Horv\'ath for useful
discussions. This work is partly supported by Slovak Grant
Agencies APVT-51-052702, VEGA 1/2009/05, and by EU through RTN
Spintronics (HPRN-CT-2000-000302).

\end{document}